\documentstyle[12pt]{article}
\newcommand{\tdot}[1]{\stackrel{\bf \,...}{\textstyle #1}}
\def\fs{\hbox{$.\!\!^{\rm s}$}}
\def\farcs{\hbox{$.\!\!^{\prime\prime}$}}
\input psfig
\makeatletter
\def\section{\@startsection {section}{1}{\z@}{2.3ex plus 1ex minus
.2ex}{1.5ex plus .2ex}{\normalsize\bf}}
\def\bibitem[#1<#2>]#3{\par\hangindent=10mm\hangafter=1
    \if@filesw
	{\def\protect##1{\string ##1\space}
	\immediate\write\@auxout{\string\lbibcite{#3}{#1}{#2}}}
    \fi}
\def\lbibcite#1#2#3{\global\@namedef{b@#1}{#2}
                    \global\@namedef{by@#1}{#3}}
\def\@bcite#1#2{({#1\if@tempswa , #2\fi})}
\def\@pcite#1#2{#1\if@tempswa , #2\fi}
\def\@brcite{\ (}
\def\@ebrcite{)}
\def\@fmttop#1{\@fmtcite[\csname #1\endcsname]}
\def\@citex[#1]#2{\if@filesw\immediate\write\@auxout{\string\citation{#2}}\fi
  \def\@citea{}\@cite{\@for\@citeb:=#2\do
    {\@citea\def\@citea{;\ }\@ifundefined
       {b@\@citeb}{{\bf ?}\@warning
       {Citation `\@citeb' on page \thepage \space undefined}}
{\csname b@\@citeb\endcsname\@citebro\csname by@\@citeb\endcsname\@citebrc}}}
{#1}}
\def\cite{\@ifnextchar [{\let\@citebro=\space
                          \let\@citebrc=\relax
                          \let\@cite=\@bcite\@tempswatrue\ \@citex}
                        {\let\@citebro=\space
                          \let\@citebrc=\relax
                          \let\@cite=\@bcite\@tempswafalse\ \@citex[]}}
\def\pcite{\@ifnextchar [{\let\@citebro=\space
                          \let\@citebrc=\relax
                          \let\@cite=\@pcite\@tempswatrue\@citex}
                        {\let\@citebro=\space
                          \let\@citebrc=\relax
                          \let\@cite=\@pcite\@tempswafalse\@citex[]}}
\def\scite{\@ifnextchar [{\let\@citebro=\@brcite
                          \let\@citebrc=\@ebrcite
                          \let\@cite=\@pcite\@tempswatrue\@citex}
                        {\let\@citebro=\@brcite
                          \let\@citebrc=\@ebrcite
                          \let\@cite=\@pcite\@tempswafalse\@citex[]}}
\def\thebibliography#1
{\section*{References\markboth {REFERENCES}{REFERENCES}}   
   \def\newblock{\hskip .11em plus .33em minus -.07em}
   \sloppy
   \sfcode`\.=1000\relax
   \parindent=0mm\parskip3mm}

\makeatother

\begin{document}
\centerline{\large \bf Orbital Variability in the Eclipsing Pulsar Binary}
\smallskip
\centerline{\large \bf PSR~B1957+20}
\bigskip
\centerline{Z. Arzoumanian,\footnote{Joseph Henry
Laboratories and Department of Physics, Princeton University,
Princeton, NJ 08544. \\ E-mail: Internet, zaven@pulsar.princeton.edu}
A. S.  Fruchter,\footnote{\noindent Astronomy Department and Radio Astronomy
Laboratory, University of California, Berkeley, CA 94720. \\ E-mail:
Internet, asf@orestes.berkeley.edu \\ Hubble Fellow} {\sc and} J. H.
Taylor\footnote{\noindent Joseph Henry Laboratories and Department of Physics,
Princeton University. \\ E-mail: Internet, joe@pulsar.princeton.edu}}

\begin{abstract}

We have conducted timing observations of the eclipsing
millisecond binary pulsar PSR~B1957+20, extending the span of
data on this pulsar to more than five years. During this time the orbital
period of the system has varied by roughly $\Delta P_b/P_b = 1.6 \times
10^{-7}$,  changing quadratically with time and displaying an orbital 
period  second derivative $\ddot P_b = (1.43 \pm 0.08) \times
10^{-18}\,$s$^{-1}$. The previous measurement of a large negative
orbital period derivative reflected only the short-term behavior of the
system during the early observations; the orbital period derivative
is now positive and increasing rapidly. If, as we suspect, the
PSR~B1957+20 system is undergoing quasi-cyclic orbital period
variations similar to those found in other close binaries such as
Algol and RS CVn, then the $0.025\,M{_\odot}$ companion to
PSR~B1957+20 is most likely non-degenerate, convective, and
magnetically active.

\bigskip
\noindent {\em Subject headings:\/} pulsars --- binaries: evolution --- stars:
eclipsing binaries --- stars: individual (PSR~B1957+20)

\end{abstract}

\section{Introduction} 
The evolutionary links between solitary millisecond pulsars and their
presumed binary progenitor systems may involve a number of exotic
astrophysical phenomena. In their late stages of evolution, neutron
stars in low-mass X-ray and pulsar binaries may evaporate their
companions through the strength of their radiation, turning themselves
into solitary ``recycled'' pulsars \cite{acrs82,rst89a,bv91}.  In some
cases, material from the companion  may even reform to create planets
\cite{bs92,tb92,srp92}.  The discovery of PSR~B1957+20 \cite{fst88}, a
1.6 ms pulsar in orbit with a $\sim0.025\,M_{\odot}$ companion, has
provided the strongest evidence that this scenario actually occurs in
nature, and that interacting binary systems are indeed responsible for
the creation of the fastest pulsars.

In spite of the small size of its companion, the pulsar's radio signal
is eclipsed over approximately ten percent of the 9.2 hour orbit.
Excess delays of the pulses are observed for many minutes after eclipse
egress but only briefly before ingress, revealing the existence of an
ionized wind from the companion which is continuously infused with new
matter (\pcite{fbb+90}, hereafter F90) and is responsible for the radio
eclipse.  Optical observations of the companion provide estimates of
its temperature, radius and thermal timescale (\pcite{fg92} and
references therein), and show a strong modulation with orbital phase of
its optical luminosity  consistent with irradiation from the pulsar.
Radio observations show  the  electron column density in the evaporated
wind  to be small (F90; \pcite{rt91b}, hereafter RT91), a result
supported by observed transparency of the wind to unpulsed emission at
$\lambda = 20$\,cm \cite{fg92}.  However, RT91 also reported a large
(negative) orbital period derivative over the $\sim$2.5 years spanned
by their observations, seemingly implying an unexpectedly short
timescale of 30$\,$Myr for orbital decay. This conclusion was difficult
to reconcile with the low rate of mass loss suggested by the density of
the companion's wind.

We have conducted further timing observations of PSR~B1957+20 beginning
in November 1992 and spanning 9 months. Together with the earlier timing
measurements described in F90 and RT91, our data reveal orbital
evolution previously unobserved in binary systems containing a pulsar:
the orbital period derivative ($\dot P_b$) of the PSR~B1957+20 system
has changed sign and has been increasing steadily.

\section{Observations and Analysis}

Our observations of PSR~B1957+20 were carried out at the Arecibo
Observatory using the Princeton
Mark~\uppercase\expandafter{\romannumeral3} system \cite{skn+92}, the
same data acquisition system used by F90 and RT91.  The pulsar signal
was coherently de-dispersed, detected, and synchronously averaged
during integrations lasting approximately two minutes.  Accurate
integration start times were provided by a local rubidium clock,
traceable via GPS to UTC.   All of our observations were made at
frequencies in the range 426--434\,MHz; two
passbands of width 0.41\,MHz were coherently dedispersed, their
center frequencies chosen to track scintillation-induced maxima in the
signal strength.  Dual circularly polarized signals were then summed to
form a total-intensity profile of pulse strength versus rotational
phase.  Pulse times-of-arrival (TOAs) were computed by fitting observed
profiles to a low-noise template and adding the resulting phase offset
to a time near the midpoint of the integration, derived by adding an
integer number of pulse periods to the start time.  TOAs collected
during 15-minute intervals were then averaged. The entire data set,
including the measurements of F90 and RT91, consists of 253 averaged
TOAs obtained between 24 March 1988 and 12 September 1991, and 111
between 19 November 1992 and 11 June 1993.  Heavy scheduling demands on
the Arecibo telescope and a timing campaign of several new millisecond
pulsars discovered at similar right ascensions are
responsible for the one year gap in the data.   A concentrated
observing session designed to yield well-sampled, redundant coverage
of the orbit was carried out during 14--20 April 1993.

We have used a modified version of the {\sc tempo} software package
\cite{tw89} to reduce the topocentric TOAs to the solar system
barycenter established by the DE200 solar system ephemeris
\cite{sta82}, and perform a multi-parameter fit to the spin,
astrometric and orbital parameters of the pulsar by minimizing the sum
of squares of the differences between predicted and observed TOAs.
Parameters of the model include the pulsar's rotational phase,
frequency and as many as six frequency derivatives (see below),
dispersion measure, position and proper motion, the projected
semi-major axis of its orbit, its orbital period and phase, two
derivatives of the orbital period, and optionally, orbital
eccentricity, angle of periastron and semi-major axis rate of change.
In all fits, pulse arrival times between orbital phases 0.19 and 0.39
(which we refer to as the ``timing eclipse'') were given zero weight;
within this region, the pulses are either eclipsed or unpredictably
delayed as they travel through the companion's wind (see F90, RT91).

The residuals from our multi-parameter  fits display long-term
unmodeled fluctuations which appear as correlated noise in the pulse
arrival times (see Figure~1).  The magnitude of these residuals is
consistent with changes in the dispersion measure of the pulsar
expected from its motion through the interstellar medium (see, e.g.,
\pcite{bhvf93}), and we believe this to be the most likely explanation
of the noise. These residuals could also be due, however, to changes in
ablated material surrounding the pulsar, or perhaps rotational
irregularities in the neutron star itself. (The existence of ``timing
noise'' in millisecond pulsars is discussed in \pcite{ktr94}.) Since
least-squares parameter estimation is unreliable when na\"{\i}vely
applied in the presence of such noise, in addition to fitting the
average deterministic spin-down behavior of the pulsar, higher-order
rotational derivatives $d^2\nu/dt^2$, $d^3\nu/dt^3$, \dots, where
$\nu=1/P$, were introduced as free parameters in order to model and
absorb the observed drifts in pulse phase. The astrometric and orbital
parameters obtained from such a fit were adopted as the best unbiased
values, and these appear in Table~1.  Since systematic parameter biases
introduced by timing noise are much larger than the formal
uncertainties in these fits, we carried out a number of fits in which
the modeled timing noise was replaced by simulated noise; we used the
resulting distributions of parameter values to obtain 1$\sigma$
estimates of the uncertainty associated with each parameter.  The
simulated noise was constructed to be similar in spectral content to
the modeled noise (represented by the high-order spin frequency
derivatives) but with arbitrary phases.  Finally, to obtain the
deterministic spin parameters of the pulsar (pulse phase, frequency,
and frequency first derivative) and to best display the timing noise,
these parameters were fit with the celestial and orbital parameters
held fixed at their adopted values. The resulting residuals are plotted
versus date and orbital phase in Figure~1.  Note that much of the power
in the ``reddest'' components of the timing noise spectrum is absorbed
in our fit for the average spin period and period derivative.  These
quantities can therefore be expected to differ, in fits which span
different epochs, by more than their measurement uncertainties. We have
not attempted to correct for any parameter bias in the global spin
period and its derivative, and so quote statistical uncertainties only
for these quantities.  Because the orbital period is some three orders
of magnitude smaller than the timescale of the timing noise, we believe
that the fitted orbital parameters are not significantly contaminated
by its effects.

The orbital eccentricity was held fixed at zero in all of our preferred
timing solutions, and $T_0$ in Table~1 is a time near the center of the
data set at which the pulsar crossed the ascending node, defined as the
zero of orbital phase.  A formal solution for the orbital eccentricity
$e$ and angle of periastron $\omega$ is in fact possible, but we prefer
to quote the results of such a fit as an upper limit to the
eccentricity, since irregular sampling in orbital phase and an angle of
periastron near eclipse egress lend little credence to the formal
solution.  In most instances and especially in the case of the pulsar's
astrometric parameters, the best-fit values presented in Table~1
improve significantly upon previous measurements.  We see no
substantial change in either the duration of eclipse or the magnitude
of excess propagation delays in comparisons of our recent observations
with the results of RT91, although these phenomena remain highly
variable from one observation to the next.  Further monitoring of
PSR~B1957+20 will add to the handful of eclipse events in our data set
and may eventually constrain any changes to the eclipsing medium.

\section{Results}

As a check of the orbital behavior implied by our global fit, we
divided our data into five non-overlapping subsets, each containing
about one year of pulse arrival times. These were individually fit for
$P$, $\dot P$, one or two additional rotational derivatives where
necessary, and two orbital parameters, $T_0$ and $P_b$, at an epoch
near the center of each subset while the pulsar's position on the sky,
determined from the proper motion,  was held fixed.  The resulting
fractional orbital period changes are plotted in Figure~2a.  A varying
gravitational acceleration due to a second companion in a distant orbit
or large fluctuations in dispersion measure could, in principle,
produce changes in the apparent orbital period similar to those seen
here. They would, however, also affect our measurement of the pulsar
spin period, so that $\Delta P_b/P_b = \Delta P/P$.  As Figure~2b
shows, the spin period displays no such variation, at a level nearly
five orders of magnitude less than that required.  Apparent variations
in orbital period due to such dynamical effects can therefore be ruled
out. While both the transverse Doppler shift due to the pulsar's proper
motion and differential Galactic acceleration can introduce a spurious,
essentially constant $\dot P_b$ \cite{dt91}, we note that for
PSR~B1957+20 they together contribute to the observed orbital period
derivative at a level comparable to the uncertainty in our measurement
of $\dot P_b$.

Figure~3 displays the orbital phase shifts detected in the PSR~B1957+20
system over more than 5 years.  The points plotted in this ``observed
minus computed'' diagram are deviations from the orbital
ephemeris\footnote{The values of $P_b$ and its derivatives compiled in
Table 1 reflect a Taylor expansion of the orbital period about the
center of the data span; by contrast, the value $P_b=33001.9162448\,$s
in Equation (1) is roughly the {\em average} orbital period over the same
span. The constant 86400 is the number of seconds in one day.}
\begin{equation} 
\phi_c = \frac{86400}{33001.9162448} [ t({\rm MJD}) - T_0],
\end{equation} 
derived from our pulse arrival-time measurements.  To obtain these
differences, estimated TOAs were computed using only the best-fit
astrometric and spin parameters (including additional frequency
derivatives) and  subtracted from the observed values.  Non-zero
residuals were thus assumed to be entirely due to orbital phase deviations
from the pre-fit model.  Residuals obtained during timing eclipse and
within 0.06 of orbital phase 0.75 (inferior conjunction) were omitted
since small random fluctuations at these phases result in large
apparent orbital phase shifts. The phase shifts derived from single
integrations were then combined, with relative weights proportional to
the cosine of the orbital phase, to form daily averages.  The results
of this process are plotted in Figure~3; error bars represent the RMS
deviation from the mean of each daily average.  The dominant cubic
trend corresponds to the value of $\ddot P_b$ derived from the global
fit.  Note that a decreasing slope (for example, from positive to
negative after 1990) in the phase shifts of Figure~3 indicates a
decrease in orbital frequency, or an increase in orbital period.  This
convention differs in sign from O$-$C diagrams derived from eclipse
timings: we measure the difference in orbital phase at a given instant
in time, while eclipse timing yields a difference in the time of
passage through a specified orbital phase.

\section{Discussion}

Although our observations reveal that the orbital period derivative
discovered by RT91 is not constant and thus provides little direct
information on the ultimate fate of the companion of PSR B1957+20, we
cannot entirely rule out the most popular explanation of that first
measurement, that the orbital period changes are caused by substantial
mass loss \cite{bs92,eic92,bt93,mfkr94}. Nonetheless, we find this
proposition unlikely for a number of reasons. In order to transport
sufficient angular momentum to produce the observed orbital period
variations, the wind density must be several orders of magnitude higher
than indicated by the electron density along the line of sight.
Therefore, the bulk of the escaping material must either be hidden by
the orbit's inclination or be overwhelmingly neutral; the latter
explanation seems particularly unlikely given the large systemic escape
velocity and the intensity of the pulsar radiation. Furthermore, to
match our present results the angular momentum carried by the ablated
matter must have varied smoothly over the past five years while
doubling in magnitude and changing sign.

We believe a more natural explanation is that PSR B1957+20 undergoes
small quasi-periodic oscillations in orbital period.
Orbital variations comparable in magnitude to those witnessed here
are fairly common in short-period binaries
containing a low-mass main-sequence star and have been well-studied in
Algol and RS~CVn systems \cite{sod80,hal89,war88}. In such binaries,
rotation of the main-sequence star is likely to be tidally locked to
the orbital period; as a result, the ratio of rotational
timescale to convective timescale, the star's ``Rossby number,''
is less than one. These stars are generally magnetically active and
display substantial chromospheric activity, radio and x-ray flares and
stellar winds $10^2$--$10^4$ times stronger than slowly rotating stars
of similar spectral class \cite{pl93,sim90}. The rapid rotation appears
to maintain a magnetic dynamo which not only creates an energetic
stellar atmosphere, but also distorts the star sufficiently to
alter its gravitational quadrupole moment and, in turn, the 
orbital period \cite{app92}.

If the companion to PSR B1957+20 is bloated and at least partially
non-degenerate, as optical observations appear to imply
\cite{arc92,fg92}, then the stellar atmosphere should be convective and
have an overturn timescale far in excess of the 9.2 hour binary
period.  The Rossby number of the companion would then be less than
one and, like the binaries discussed above, the system might be
expected to display orbital period variations and the companion a
strong stellar wind, even in the absence of a pulsar primary. Although
the companion's external magnetic field is already constrained by
Faraday delay measurements (less than a few gauss parallel to the line
of sight at the edges of the eclipse region), a much stronger,
toroidal, subsurface field could remain undetected by these
observations.  If this hypothesis is true, the truly peculiar aspect of
the system is not the activity of the companion, but rather its
non-degeneracy, for this star is far too light to be burning hydrogen.
Either the present irradiation by the pulsar must be responsible for
the swollen state of this object, perhaps through a mechanism similar
to that proposed by \scite{pod91} for low-mass X-ray binaries, or
evaporation of most of the companion's mass must have been sufficiently
recent that it has not yet had time to shrink to degeneracy.

We suspect that rotationally-induced magnetic activity not only
explains much of the behavior of the PSR B1957+20 system, but may also be
important in understanding observations of two other eclipsing pulsars
in short-period binaries, PSRs B1744$-$24A \cite{lmd+90} and B1718$-$19
\cite{lbhb93}. The companions of these pulsars are most likely low-mass
main-sequence dwarfs, and in the case of PSR B1744$-$24A, the companion
should nearly fill its Roche lobe. Both of these objects display
evidence of excess material surrounding the entire binary, PSR B1744$-$24A
through prolonged ``anomalous'' eclipses and pulse arrival delays
\cite{lmd+90,nttf90} and PSR B1718$-$19 through an inverted radio
spectrum below 600\,MHz. In each case, however, the energy density
of the pulsar radiation impinging on the companion is far less than that
to which the companion of PSR B1957+20 is exposed. (The observed
spin-down rate of PSR B1744$-$24 is certainly contaminated by the
system's acceleration in the gravitational potential of its 
cluster, but the pulsar flux at the companion can be estimated
by assuming an intrinsic period derivative similar to that of other
millisecond pulsars, \pcite{nt92}.) While pulsar irradiation would seem
incapable of expelling the observed material from the companion
surfaces, a rotationally powered wind, such as those found in RS~CVn 
systems,
could be sufficiently strong to produce the eclipses (a point that has
been made independently by \pcite{wp93} for PSR B1718$-$19) and could
explain the seemingly ``episodic'' nature of the anomalous
eclipses in PSR B1744$-$24A \cite{lmd+90,nt92}. One might expect to see
orbital period variations in these other two eclipsing systems, but the
low flux densities and long spin periods of these pulsars may make the
required timing accuracy difficult to obtain.

\medskip 
M. F. Ryba and D. R. Stinebring built observing hardware and obtained
some of the data that made this project possible. We are in their
debt.  We are also grateful to F. Camilo and A.  V\'{a}zquez for
observing assistance, and to D. J. Nice, B.  Paczy\'{n}ski, and C.
Thompson for helpful discussions.  The Arecibo Observatory is part of
the National Astronomy and Ionosphere Center, operated by Cornell
University under cooperative agreement with the National Science
Foundation. ASF was supported by a Hubble Fellowship awarded by NASA
through the Space Telescope Science Institute.

\newpage
\begin{table}
\begin{center}
\caption{Astrometric, Spin, and Orbital Parameters of PSR~B1957+20.}
\begin{tabular}{ll}
\hline
\hline
Right ascension, $\alpha$ (J2000)$^a$ \dotfill & $19^{\rm h}\,59^{\rm m}\,
36\fs76988(5)$ \\
Declination, $\delta$ (J2000) \dotfill & $20^\circ\,48'\,15\farcs1222(6)$ \\
$\mu_{\alpha}\,({\rm mas}\;{\rm yr}^{-1})$ \dotfill & $-16.0\pm0.5$ \\
$\mu_{\delta}\,({\rm mas}\;{\rm yr}^{-1})$ \dotfill & $-25.8\pm0.6$ \\
Period, $P$ (ms) \dotfill & 1.60740168480632(3) \\
Period derivative, $\dot P$ ($10^{-20}$) \dotfill & 1.68515(9) \\
$\ddot P$ ($10^{-31}\,{\rm s}^{-1}$) \dotfill & $1.4\pm0.4$ \\
Epoch (MJD) \dotfill & 48196.0 \\
Dispersion measure, DM (cm$^{-3}\;$pc) \dots & 29.1168(7) \\
Projected semi-major axis, $x$ (lt-s) \dotfill & 0.0892253(6) \\
Eccentricity, $e$ \dotfill & $< 4 \times 10^{-5}$ \\ 
Epoch of ascending node, $T_0$ (MJD) \dotfill & 48196.0635242(6) \\
Orbital period, $P_b$ (s) \dotfill & 33001.91484(8) \\
$\dot P_b$ (10$^{-11}$) \dotfill & 1.47$\pm$0.08 \\
$\ddot P_b$ (10$^{-18}\,{\rm s}^{-1}$) \dotfill & 1.43$\pm$0.08 \\
$|\tdot P_b|$ (10$^{-26}\,{\rm s}^{-2})$ \dotfill & $< 3$ \\
$|\dot x|$ (10$^{-14}$) \dotfill & $< 3$ \\
\hline
\hline
\end{tabular}
\end{center} 
\noindent $^a$Coordinates are given in the J2000 reference frame of
the DE200 solar system ephemeris. Figures in parentheses are
uncertainties in the last digits quoted.  
\end{table}

\newpage
\section*{Figure Captions}

\bigskip
\bigskip
Figure 1: Post-fit residuals of PSR~B1957+20 plotted versus date
and orbital phase. The typical uncertainty in the pulse arrival times,
a few microseconds, is shown near the upper left in a).

\bigskip
\bigskip
Figure 2: a) Fractional orbital period changes in the PSR~B1957+20
system versus date. The overall variation in $P_b$ spans about 5\,ms.
The solid line curve corresponds to the values of $P_b$ and its
derivatives listed in Table~1. b) Fractional pulse period changes
versus date. A correction has been made for the global
spin-down rate given by $\dot P$ in Table~1. Error bars reflect
only the statistical uncertainties from the individual fits and do not
include the possible effects of the long-term timing noise.

\bigskip
\bigskip
Figure 3: Orbital phase shifts (observed minus computed) in the
PSR~B1957+20 system. Note that the sign convention employed here is the
opposite of that in the orbital phase shift plot of RT91, Figure 3. The
solid line is derived from Equation (1) and the orbital information
given in Table 1.

\clearpage
\begin{figure}
\centerline{\psfig{figure=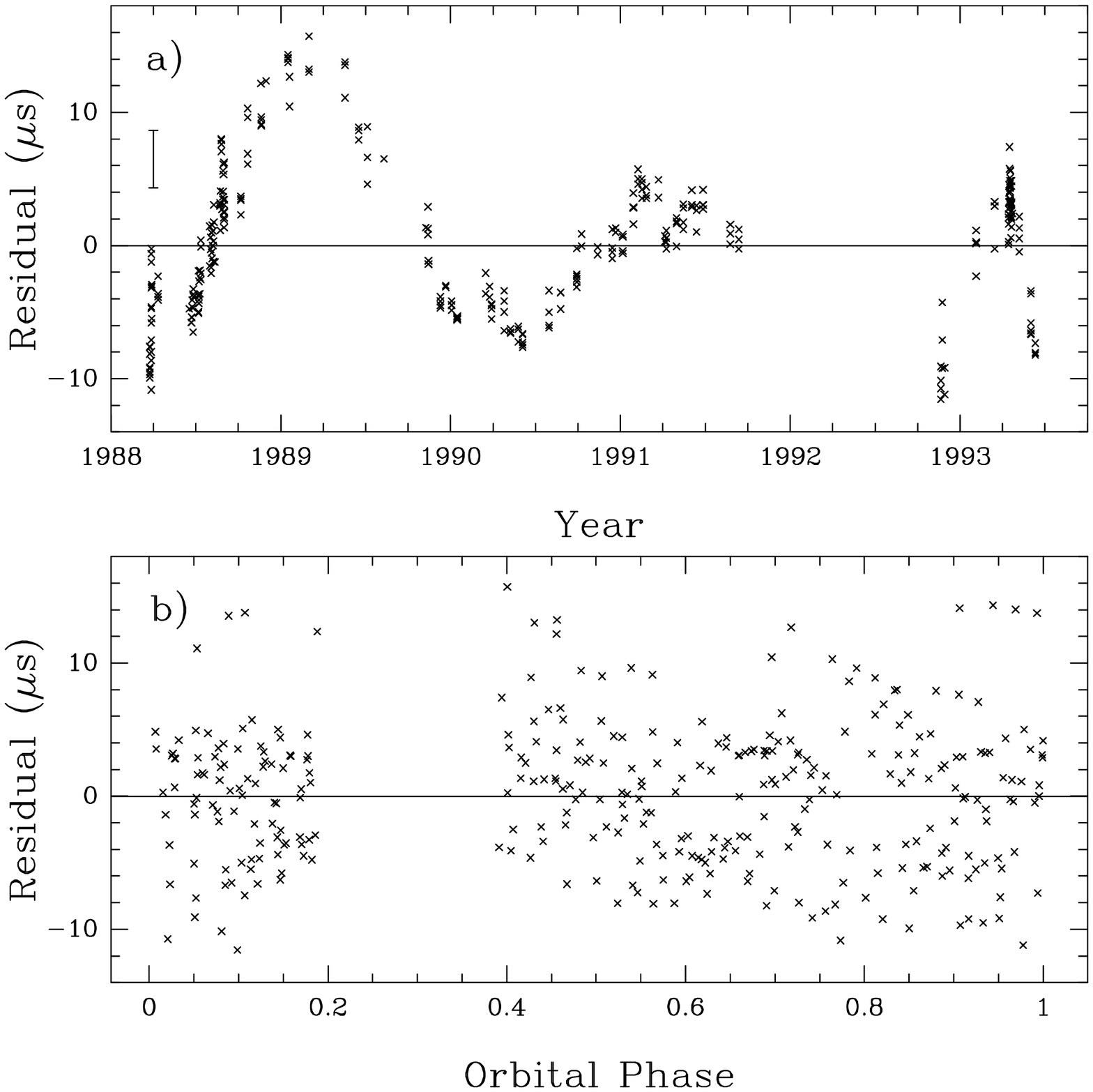,width=8in}}
\end{figure}
 
\begin{figure}
\centerline{\psfig{figure=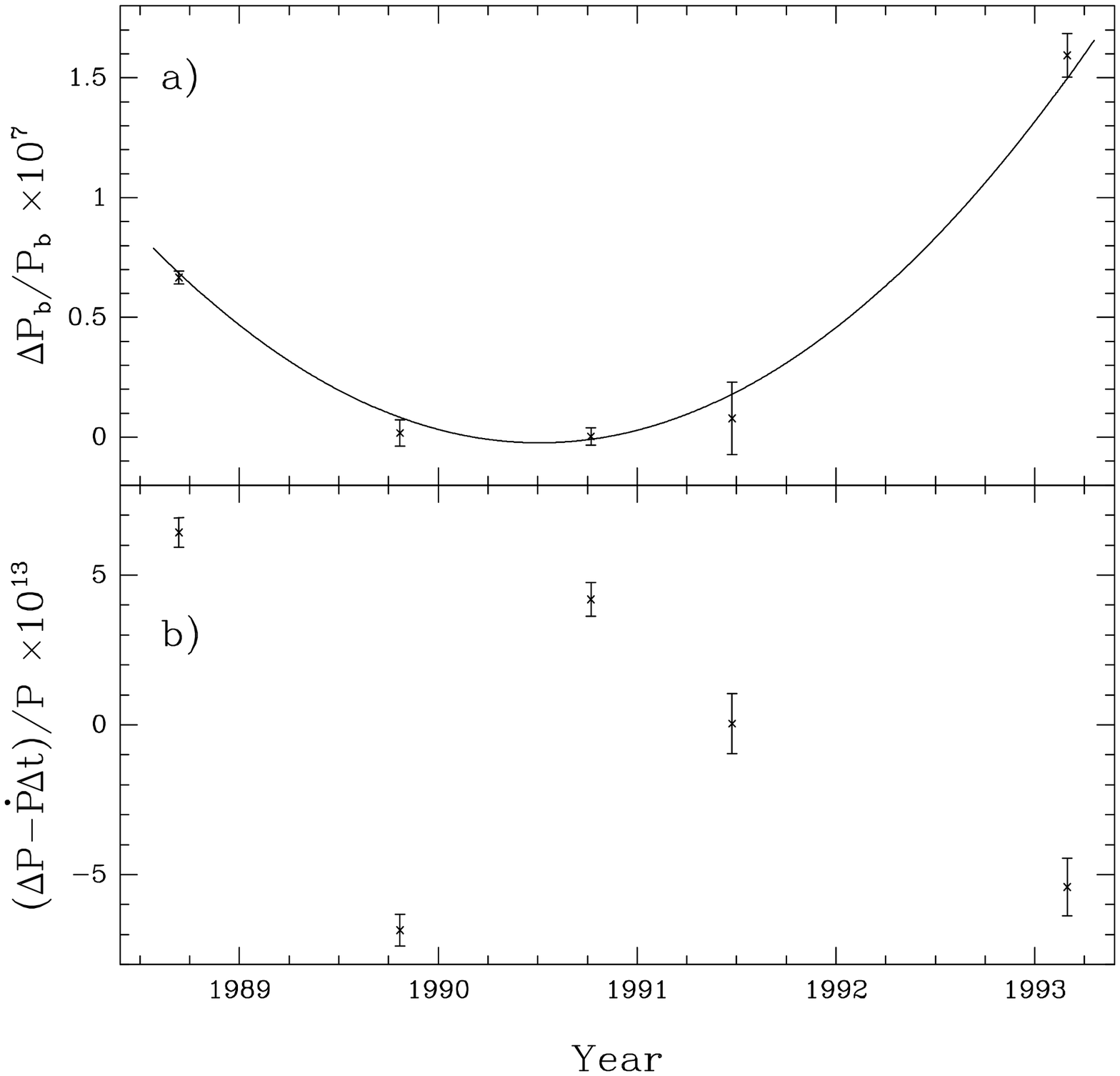,width=8in}}
\end{figure}
 
\begin{figure}
\centerline{\psfig{figure=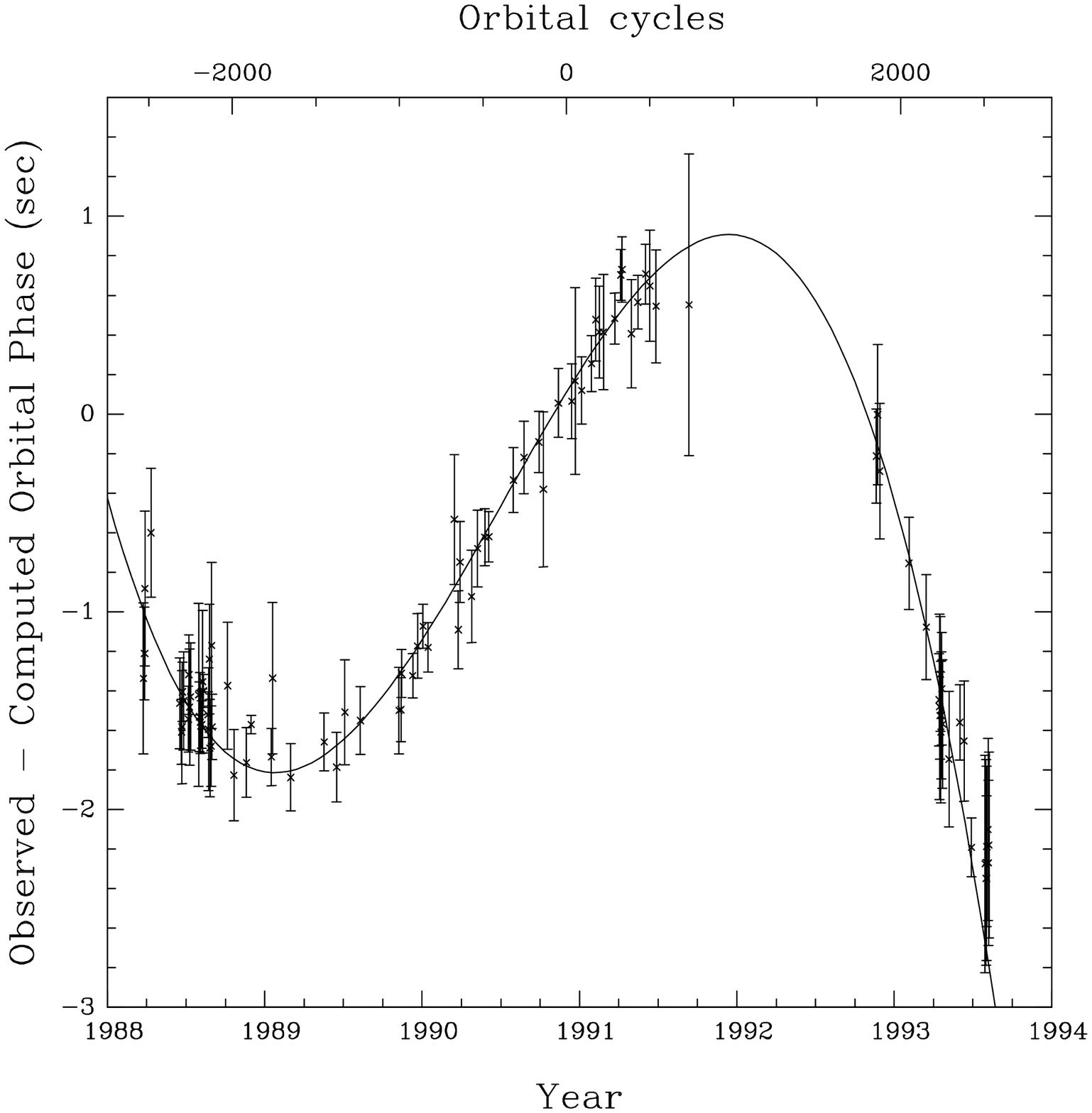,width=8in}}
\end{figure}


\newpage
\begin{thebibliography}{Bhattacharya \& van~denHeuvel<1991>}

\bibitem[Aldcroft, Romani \& Cordes<1992>]{arc92}
Aldcroft, T.~L., Romani, R.~W.  \& Cordes  J.~M.  1992,
\newblock ApJ, { 400}, 638

\bibitem[Alpar {\it et~al.}<1982>]{acrs82}
Alpar, M.~A., Cheng, A.~F., Ruderman, M.~A.  \& Shaham  J.  1982,
\newblock Nature, { 300}, 728

\bibitem[Applegate<1992>]{app92}
Applegate  J.~H.  1992,
\newblock ApJ, { 385}, 621

\bibitem[Backer {\it et~al.}<1993>]{bhvf93}
Backer, D.~C., Hama, S., Van~Hook, S.  \& Foster  R.~S.  1993,
\newblock ApJ, { 404}, 636

\bibitem[Banit \& Shaham<1992>]{bs92}
Banit, M. \& Shaham  J.  1992,
\newblock ApJ, { 388}, L19

\bibitem[Bhattacharya \& van~den~Heuvel<1991>]{bv91}
Bhattacharya, D. \& van~den Heuvel, E. P.~J.  1991,
\newblock Phys.\,Rep.\,, { 203}, 1

\bibitem[Brookshaw \& Tavani<1993>]{bt93}
Brookshaw, L. \& Tavani  M.  1993,
\newblock ApJ, { 410}, 719

\bibitem[Damour \& Taylor<1991>]{dt91}
Damour, T. \& Taylor  J.~H.  1991,
\newblock ApJ, { 366}, 501

\bibitem[Eichler<1992>]{eic92}
Eichler  D.  1992,
\newblock MNRAS, { 254}, 11P

\bibitem[Fruchter {\it et~al.}<1990>]{fbb+90}
Fruchter, A.~S., Berman, G., Bower, G., Convery, M., Goss, W.~M., Hankins,
  T.~H., Klein, J.~R., Nice, D.~J., Ryba, M.~F., Stinebring, D.~R., Taylor,
  J.~H., Thorsett, S.~E.  \& Weisberg  J.~M.  1990,
\newblock ApJ, { 351}, 642

\bibitem[Fruchter \& Goss<1992>]{fg92}
Fruchter, A.~S. \& Goss  W.~M.  1992,
\newblock ApJ, { 384}, L47

\bibitem[Fruchter, Stinebring \& Taylor<1988>]{fst88}
Fruchter, A.~S., Stinebring, D.~R.  \& Taylor  J.~H.  1988,
\newblock Nature, { 333}, 237

\bibitem[Hall<1989>]{hal89}
Hall  D.~S.  1989,
\newblock Space Sci. Rev., { 50}, 219

\bibitem[Kaspi, Taylor \& Ryba<1994>]{ktr94}
Kaspi, V.~M., Taylor, J.~H.  \& Ryba  M.  1994,
\newblock ApJ, { in press}

\bibitem[Lyne {\it et~al.}<1993>]{lbhb93}
Lyne, A.~G., Biggs, J.~D., Harrison, P.~A.  \& Bailes  M.  1993,
\newblock Nature, { 361}, 47

\bibitem[Lyne {\it et~al.}<1990>]{lmd+90}
Lyne, A.~G., Manchester, R.~N., D'Amico, N., Staveley-Smith, L., Johnston, S.,
  Lim, J., Fruchter, A.~S., Goss, W.~M.  \& Frail  D.  1990,
\newblock Nature, { 347}, 650

\bibitem[McCormick {\it et~al.}<1994>]{mfkr94}
McCormick, P.~J., Frank, J., King, A.~R.  \& Rajasekhar  A.  1994,
\newblock ApJ, { submitted}

\bibitem[Nice \& Thorsett<1992>]{nt92}
Nice, D.~J. \& Thorsett  S.~E.  1992,
\newblock ApJ, { 397}, 249

\bibitem[Nice {\it et~al.}<1990>]{nttf90}
Nice, D.~J., Thorsett, S.~E., Taylor, J.~H.  \& Fruchter  A.~S.  1990,
\newblock ApJ, { 361}, L61

\bibitem[Pasquini \& Lindgren<1993>]{pl93}
Pasquini, L. \& Lindgren  H.  1993,
\newblock A\&A, { submitted}

\bibitem[Podsiadlowski<1991>]{pod91}
Podsiadlowski  P.  1991,
\newblock Nature, { 350}, 136

\bibitem[Ruderman, Shaham \& Tavani<1989>]{rst89a}
Ruderman, M., Shaham, J.  \& Tavani  M.  1989,
\newblock ApJ, { 336}, 507

\bibitem[Ryba \& Taylor<1991>]{rt91b}
Ryba, M.~F. \& Taylor  J.~H.  1991,
\newblock ApJ, { 380}, 557

\bibitem[Simon<1990>]{sim90}
Simon  T.  1990,
\newblock ApJ, { 359}, L51

\bibitem[S\"oderhjelm<1980>]{sod80}
S\"oderhjelm  S.  1980,
\newblock A\&A, { 89}, 100

\bibitem[Standish<1982>]{sta82}
Standish  E.~M.  1982,
\newblock A\&A, { 114}, 297

\bibitem[Stevens, Rees \& Podsiadlowski<1992>]{srp92}
Stevens, I.~R., Rees, M.~J.  \& Podsiadlowski  P.  1992,
\newblock MNRAS, { 254}, 19P

\bibitem[Stinebring {\it et~al.}<1992>]{skn+92}
Stinebring, D.~R., Kaspi, V.~M., Nice, D.~J., Ryba, M.~F., Taylor, J.~H.,
  Thorsett, S.~E.  \& Hankins  T.~H.  1992,
\newblock Rev. Sci. Instrum., { 63}, 3551

\bibitem[Tavani \& Brookshaw<1992>]{tb92}
Tavani, M. \& Brookshaw  L.  1992,
\newblock Nature, { 356}, 320

\bibitem[Taylor \& Weisberg<1989>]{tw89}
Taylor, J.~H. \& Weisberg  J.~M.  1989,
\newblock ApJ, { 345}, 434

\bibitem[Warner<1988>]{war88}
Warner  B.  1988,
\newblock Nature, { 336}, 129

\bibitem[Wijers \& Paczy\'{n}ski<1993>]{wp93}
Wijers, R. A. M.~J. \& Paczy\'{n}ski  B.  1993,
\newblock ApJ, { 415}, L115

\end{thebibliography}
\end{document}